# Title: Quantifying the high early solar cosmic ray flux with cosmogenic neon isotopes in refractory minerals

## Short title: an active young Sun beyond expectation

Xin Yang[1,2], Fred J. Ciesla[1], and Philipp R. Heck[1,2]

[1]Department of the Geophysical Sciences and Chicago Center for Cosmochemistry, The University of Chicago, Chicago, IL 60637, USA

[2]Robert A. Pritzker Center for Meteoritics and Polar Studies, Negaunee Integrative Research Center, Field Museum of Natural History, Chicago, IL, USA

**Abstract:** An enhancement in the activity of the early young Sun resulting in a high charged particle flux has been invoked to explain excesses in spallation-induced nuclides in primitive planetary materials. Astronomical observations of energetic outbursts of young stellar objects (YSOs) also support the idea of an active young Sun. However, the early solar cosmic-ray (SCR) flux has not been well constrained. Here we use measured concentrations of SCR-produced nuclides that formed and are preserved in meteoritical hibonite and spinel, some of the Solar System's oldest solids, and physical models for dust transport in the early protoplanetary disk to determine the magnitude of the early SCR flux. We focus our attention on cosmogenic neon which cannot have been inherited from precursors and can only be produced in situ in solids. Our modeled effective exposure time to SCRs for these solids is very short on the order of years. This indicates the young Sun's SCR flux recorded in refractory mineral hibonite was up to ~7 orders of magnitude higher than the contemporary level. Our flux estimate is consistent with the >$10^5$× enhanced flux inferred from astronomical observations of greatly enhanced flare activities of YSOs.

# 1. INTRODUCTION

The activity of the early young Sun consisted of complex dynamic processes such as magnetic reconnection events that produced intense flares, released bursts of energetic charged particles and X-rays, and affected the early evolution of the Solar System (Low 1996; Weiss & Tobias 2000; Ribas et al. 2005). Though the present-day activity is well-quantified, and a highly active young Sun is suggested, the early solar activity is poorly constrained, which constitutes a missing piece in the history of the Sun (Bekaert et al. 2021). Astronomical observations of strongly variable, high-luminosity X-ray emission from young stellar objects (YSOs) are evidence for a high flare activity of young solar mass stars associated with at least $10^5$-fold enhancement in the Solar cosmic ray (SCR) flux compared to contemporary levels (Feigelson et al. 2002; Feigelson 2010; Getman & Feigelson 2021). These observations, together with evidence of an active protosun from charged particle spallation records preserved in meteorites, indicate an early solar activity similar to astronomically observed YSOs. Excesses in the isotopes $^3$He, $^6$Li, $^7$Be, $^{10}$Be, $^{21}$Ne, $^{36}$Cl, $^{41}$Ca, and $^{50}$V detected in meteoritic refractory minerals are attributed to spallogenic production from a high flux of charged particle from energetic solar flares (Gounelle et al. 2001; Liu et al. 2010; Dauphas & Chaussidon 2011; Trappitsch & Ciesla 2015; Sossi et al. 2017). Understanding the production of these nuclides thus requires understanding the activity level of the young Sun and the manner in which solids were exposed to SCRs before incorporation into planetesimals. Also, while the evidence for an active young Sun is compelling, the knowledge of the average exposure time of dust grains to SCRs and a more quantitative estimate of the early solar activity is desirable. Here, we address these knowledge gaps with physical models and isotopic data from early condensates from the solar nebula that are preserved unaltered in Murchison, a primitive meteorite.

Firstly, cosmogenic $^{21}$Ne ($^{21}$Ne$_{cos}$) excess, which was produced via spallation, correlates with the presence of solar flare heavy ion tracks; this was first observed in a small fraction of meteoritic olivine grains, which was interpreted as pre-compaction irradiation at the parent body surface (a few centimeters) to a several orders of magnitude higher SCR flux for < one Myr in the early Solar System (Caffee et al. 1987; Hohenberg et al. 1990). However, an alternative irradiation scenario without an enhanced SCR flux could not be excluded to explain the $^{21}$Ne$_{cos}$ record in olivine grains. It was argued that the solids were exposed repeatedly to galactic cosmic rays (GCRs) over long time periods

(hundreds of Myr) during their residence on the parent body's layer of loosely consolidated surface materials of a few meters depth, the regolith, which gets overturned by impacts, known as regolith gardening (Wieler et al. 2000; Roth et al. 2011; Riebe et al. 2017). Whether an enhanced SCR flux left its mark in meteoritic solids was not resolved until Kööp et al. (2018) found excess cosmogenic $^{21}Ne_{cos}$ in more than 78% of platy hibonite crystals (PLACs) from the carbonaceous CM2 chondrite Murchison and in only ~7% of spinel-rich inclusions in the same Murchison meteorite specimen. PLACs and spinel-rich inclusions in Murchison are both refractory oxide mineral grains and are believed to be the first condensates in the solar nebula. For simplicity, we will call them hibonite and spinel hereafter. The bimodal distribution of cosmogenic Ne in spinels and hibonites separated from the same meteorite cannot be explained by exposure to cosmic rays after parent body accretion. While the $^{21}Ne_{cos}$ record in spinel is similar to the one in olivine grains and other meteorite components (Caffee et al. 1987; Hohenberg et al. 1990; Roth et al. 2011; Riebe et al. 2017), the hibonites display a strikingly different irradiation record. That difference requires that the hibonites were irradiated as individual particles before they accreted to the Murchison parent asteroid (Kööp, et al. 2018). The Ne isotopic composition of hibonites ($^{21}Ne/^{22}Ne<0.8$) indicates SCR-produced Ne, as the GCR-produced Ne has a higher $^{21}Ne/^{22}Ne$ ratio ($^{21}Ne/^{22}Ne\sim0.9$). Therefore, the cosmogenic neon excess in hibonites from Murchison must be due to the direct exposure to an early SCR flux as free-floating grains and not due to exposure after the parent body accretion. Our best explanation for $^{21}Ne_{cos}$ record in spinels is the same as proposed in previous work, the exposure on the parent body to SCRs or GCRs. Alternatively, they could have been irradiated as free-floating particles before accretion to the parent body by a lower SCR flux compared to the flux that the hibonites were exposed to.

Secondly, refractory minerals in primitive meteorites are believed to be the first solids formed in the Solar System, originating in the inner edge of the nebula and then transported to the outer Solar System where they were accreted by planetesimals (Wood 2004). Various hypotheses have been proposed to explain the outward transport of particles in the solar nebula, including gravitational torques in a marginally-unstable disk (Boss 2008), photophoresis (Wurm et al. 2010), radiation pressure (Vinković 2009), disk winds (Salmeron & Ireland 2012; Van Kooten et al. 2016), diffusive transport through the disk (Cuzzi et al. 2003; Ciesla 2007; Ciesla 2010a; Jacquet et al. 2011; Desch et al. 2018), and the X-wind (Shu et al. 1996). We separate the different transport mechanisms into two categories: "above the disk" and "within the disk" transport (Fig. 1).

The exposure of grains to SCRs occurs outside the disk in the "above the disk" model, whereas exposure would be limited to the very upper layers of the disk in the "within the disk" models due to the relatively short stopping distances over which the SCR-particles are attenuated in hydrogen-rich gas. Here we investigate which of these two classes of models provide the best explanation for the spallation features observed, while noting it is conceivable that both processes have contributed to dust grain transport in the early Solar System.

Thirdly, Ne is volatile, and the host minerals have low closure temperature. That means, (1) the $^{21}Ne_{cos}$ excess observed in refractory minerals must have been accumulated in a relatively cool environment where the accumulation rate surpasses the rate of thermal loss; (2) The acquisition of $^{21}Ne_{cos}$ by refractory minerals can only occur during their outward transport to the outer regions of the Solar System or thereafter, rather than during their formation in a high-temperature region. By employing a modeling approach to assess the thermal history of solid materials within the Solar System, we are able to investigate transport mechanisms that lead to retained $^{21}Ne_{cos}$ and propose a plausible explanation for the $^{21}Ne_{cos}$ record in refractory minerals.

In this paper, we aim to determine the flux of SCRs emitted by the young Sun by quantifying the levels needed to produce the observed $^{21}Ne_{cos}$ in refractory meteoritic minerals as they are exposed in the different transport models. That is, with the known concentrations of $^{21}Ne_{cos}$ from previous research (Kööp, et al. 2018) and the exposure time to SCRs acquired by modeling the particle movement in the early Solar System, we can calculate the time-averaged production rate of the nuclide. By comparing it to the present-day production rates (Trappitsch & Leya 2013), we can estimate the early SCR flux required to explain the irradiation record in meteoritic minerals. In our model, we assume that hibonite and spinel were transported and exposed in the same way before accretion, even though we are aware that other possibilities cannot be excluded for spinel. As such, we provide a value of a nominally enhanced SCR flux inferred from spinel cosmogenic Ne data based on this scenario.

In the next section we discuss the thermal diffusion of noble gases in minerals with varying sizes under different temperatures that offers constraints on the following sections. In section 3 we discuss different models of dust transport and the potential exposure of solids to SCRs in these models. In section 4 we calculate the effective exposure times after considering the thermal history and the intensity of the SCR flux in the early Solar System. In section 5 we analyze the overall uncertainty and compare our results to astronomical observations. Section 6 provides the conclusion.

## 2. THERMAL DIFFUSION OF NOBLE GASES IN REFRACTORY MINERALS

The loss of gas at a mineral scale is mostly attributed to thermal diffusion, and the controls on that have been well-studied (Mussett 1969; Cherniak & Watson 2013). The fractional loss from a sphere of radius *a* can be approximated by Crank (1979):

$$F_L = 1 - \frac{6}{\pi^2}\sum_{n=1}^{\infty}\frac{1}{n^2}\exp\left(-\frac{Dn^2\pi^2 t}{a^2}\right) \tag{1}$$

Where $F_L$ is fraction of gas lost, $t$ is the diffusion time, and $D$ is the diffusivity of noble gases in the host mineral, in the Arrhenius law for diffusion:

$$D = D_0 e^{-E_a/RT} \tag{2}$$

Here $D_0$ is the diffusivity at infinite temperature, $E_a$ is the activation energy, $R$ is the molar gas constant, and $T$ is the temperature. $E_a$ and $D_0$ for noble gases in materials are determined by petrological experiments (Baxter 2010); unfortunately, these two parameters for the refractory minerals of interest here have not yet been obtained. Therefore, we use data from the literature on similar minerals to approximate the diffusion rates in hibonites and spinels. First, we use He data to constrain the thermal diffusion of Ne. Since Ne diffuses more slowly than He due to its lower $D_0$ and higher $E_a$ in most minerals, these calculations for He provide an upper limit for the thermal loss of Ne (Baxter 2010; Mukhopadhyay & Parai 2019). Secondly, Al-Mg-spinel has not been studied for such a purpose, and we use magnetite, another spinel group mineral with the same crystal structure, to estimate the diffusivity. Thirdly, the diffusion coefficients for hibonite have not been determined, but it is known to have higher gas retentivity than spinel but lower than SiC (Kööp, et al. 2018; Heck et al. 2020). Thus, model predictions for the thermal loss of He and Ne from magnetite and SiC would serve as the upper and lower limits, respectively, for thermal loss from hibonite.

Using the available diffusion parameters of SiC and magnetite (Blackburn et al. 2007; Cherniak et al. 2016) we calculate He diffusive loss at different temperatures for grains with radii of 5, 15, and 50 μm. The results are plotted in Fig. 2. In the refractory minerals' forming region with T > 1570 K, spinel would lose all the He within tens of minutes, and hibonite would lose its He in less than a few hours. Also, we infer that Ne could not be retained in hibonite and spinel for more than a few days and any inherited light noble gases from precursors were lost. The results here provide constraints on the dust transport models in the next section.

## 3. DUST TRANSPORT AND THEMAL HISTORY

### 3.1. Within the Disk: Particle Diffusion and Turbulence

During the dynamic evolution of the solar nebula, materials could have been stirred and lofted to significant heights by gaseous motions in the disk due to processes associated with large-scale mass and angular momentum transport. Turbulence, if present, would offset the gravitational settling that particles would otherwise experience, allowing solids to reach high altitudes above the disk midplane. Particle tracking methods in a turbulent disk (Ciesla 2010b; Ciesla & Sandford 2012; Trappitsch & Ciesla 2015) have been used to evaluate the total time that grains may be reside at the surface of the disk and thus be irradiated by SCRs.

Trappitsch and Ciesla (2015) explored how long calcium-aluminum-rich inclusions (CAIs) particles may be exposed to SCRs within a hot, massive solar nebula of the type expected when CAIs formed. They found that at 0.3 AU, particles would have to be lofted to above 4 disk scale heights just to see the SCRs, as these energetic particles are rapidly attenuated as they move through the nebular gas. Over a period of 1 million years, *if* particles remained at this radial location, their cumulative time spent in this irradiation layer would be <100 years, and even less time at higher altitudes where SCRs would not be attenuated. However, in order to be preserved in the disk and accreted by chondritic meteorite parent bodies, CAIs cannot remain so close to the Sun for such a long period of time. Instead, the particles must be carried outward by diffusion and/or advective motions within the gas. Ciesla (2010a) showed that surviving CAIs in a viscously evolving disk would be dominated by those that formed close to the Sun in the first ~100,000 years, reducing the possible exposure timescale given above by a factor of at least 10.

Further, the residence time of small particles at a particular vertical height of a turbulent disk will mirror the distribution of the gas density; that is, particles spend more time in regions of high gas density, such as the disk midplane, and less time where gas densities are lowest, such as the disk surface (Ciesla 2010b). Thus, particles irradiated at the surface would inevitably make their way down to the disk midplane, which in a viscously evolving disk producing CAIs, would necessarily have temperatures >1500 K

(Yoneda & Grossman 1995). Even if particles were irradiated in the disk surface layers, they would be warmed upon return to the deeper interiors of the disk, and lose any Ne produced from their excursion to high altitudes due to these high temperatures. All Ne would be lost within a few days for spinel and hibonite if T > 1500 K.

In Zega et al. (2021), the outward transport of solids from the CAI formation region in a hot, young solar nebula was studied, identifying "surviving" particles that resided in the disk after 100,000 years outside of the CAI factory. The transport considered was 2D, looking at both radial and vertical motions to determine the physical conditions that each particle saw. We reinvestigated the paths of these surviving particles and found, within our model runs, that none of the survivors actually made their way to sufficiently high altitudes to be exposed to SCRs within the disk. This can be explained as described by (Ciesla 2010b, 2011), with the residency times of particles in a given region being proportional to the expected number density of particles in that same region. That is, in a turbulent protoplanetary disk, solids will have the greatest densities at the disk midplane and fall off to higher altitudes as $exp(-z^2/2H^2)$, where $z$ is the height above the disk midplane, and $H$ is the scale-height of the disk (here we assume the solids are well-coupled to the gas; if they are not their effective scale-height would be less than that of the gas and the arguments here would be even stronger). Particles will thus spend much of their time around the disk midplane, where densities are highest, and experience only short, infrequent excursions to very high altitudes (Ciesla & Sandford 2012). Primary SCRs are only able to penetrate through the top ~1 g/cm$^2$ of nebular gas, meaning that solids will only be exposed if they reach regions where all but this amount of gas exists between them and the young Sun. In young, massive disks like that expected when CAIs were forming, the total surface densities near the Sun should be very high. Taking the Gaussian distribution of density, and assuming a surface density given by $\Sigma(r)=12,000(r/1$ $AU)^{-1}$g cm$^{-2}$ (Zega et al. 2021), means that the SCR exposure layer would be the top >99.99% of the disk at 0.3 au, and thus particles, given enough time, would only be exposed to SCRs for <0.01% of their times in the disk. For a period of 100,000 years, this would amount to <10 years of exposure, but again, only if the particles remained at that location. Survival of grains, as discussed by Ciesla (2010b) and Zega et al. (2021), occurs when grains rapidly move outwards from the CAI region, meaning the amount of time spent close to the Sun is likely much less. This would result in only a small percentage of primitive solids being exposed to SCRs at the surface, which runs counter to the finding of Kööp et al. (2018) who found more than 78% of hibonites showed evidence of such irradiation. As such, we conclude that in a standard viscous disk model driven primarily by diffusion and turbulence, refractory grains are unlikely to get exposed to SCRs to produce the amount of $^{21}$Ne$_{cos}$ measured in hibonites. Whether other dynamical scenarios for "within the disk" transport in the solar nebula can yield significant exposure to SCRs while preserving the grains should be the subject of future work.

## 3.2. Above the Disk: Wind-driven Ejection

### 3.2.1. Surface of the Early Disk

In the wind-driven model, we only focus on the particle movement above the disk and not the long-term transport in the disk. For simplicity, we adopt the same disk structure as that in previous work (Trappitsch & Ciesla 2015; Zega et al. 2021) and refer the reader to details in those papers. We begin with a disk that has surface density given by $\Sigma_0$=12,000(r/1 AU)$^{-1}$ g cm$^{-2}$ from 0.1 to 100 AU. The $\Sigma_0$ is taken from Zega et al. (2021) and is higher than that in Trappitsch & Ciesla (2015) and may present an earlier solar nebula structure when refractory minerals formed.

In order to calculate the production rates of $^{21}$Ne$_{cos}$, we have to determine where particles would be exposed to SCRs while in transport above and back into the disk. Calculations of the interaction between energetic particles and the disk show that with the current SCR spectrum, the primary high-energy particles can penetrate ~1 g cm$^{-2}$ disk material without losing too much energy to trigger spallation reaction (Trappitsch & Ciesla 2015). However, the SCR particles do not enter the disk vertically or horizontally. Considering a point *P* with radial distance $r_p$ from the young Sun and the height $z_p \neq 0$ away from the midplane. We assume that the SCR particles arrive at *P* after traveling in a straight line. When the material along this path exceeds 1 g cm$^{-2}$, the height would be the surface boundary. Trappitsch and Ciesla (2015) found that the material vertically above the surface is much less than that along a straight line between the young Sun and *P*. To simplify the calculation, instead of calculating the material along the traveling path, we resolve the mass above the surface and make it equal to 0.1 g cm$^{-2}$ as an approximation. At *r* = 0.3 AU from the young Sun, this gives a surface height of 0.054 AU, consistent with the surface layer with the disk height ranging from 0.05 to 0.06 AU above the midplane, as discussed in Trappitsch and Ciesla (2015). We do not consider cosmogenic nuclide production by scattered primary SCR particles which would have to be addressed in future work on production rate models. Thus, as mentioned in the previous section our production rates are lower limits and the determined SCR fluxes are upper limits.

*3.2.2. Ejection and Travel above the Disk*

In many pre-stellar models, winds are expected to be launched nearly perpendicular to the disk, either from disk-star interface (Shu et al. 1996; Liffman 2005; Hu 2010; Salmeron & Ireland 2012) or over more extended distances. Small solids might be entrained in these winds and lofted to high altitudes, only to fall back onto the disk at larger distances. It has been suggested that only small grains, <1 μm, could be lifted by some winds (Safier 1993; Bans & Königl 2012; Herbst et al. 2021), while others have argued that particles of more than tens of micrometers could be redistributed (Shu et al. 1996; Hu 2010; Salmeron & Ireland 2012). For practical purposes, we ignore the source of the wind and assume that the refractory minerals in the size ranges we study here were propelled into ballistic trajectories away from the young Sun at various velocities and launch angles. We assume the particles to be spherical but note that the non-spherical shapes of the hibonites and spinels affect their aerodynamics during launch and entrainment in the winds.

To investigate how particles may become enriched in $^{21}Ne_{cos}$, we model the trajectory of particles after they are launched from the inner region of the disk and track how long they are exposed to the SCRs before falling below the disk surface defined in the previous section. We define the "launch" area of the particles as being above the inner edge of the disk, taken here to be 0.1 AU, where we assume the solids leave the disk and the wind. The height above the disk midplane from which particles would be launched is uncertain, and we take the region to begin at 0.0136 au, which is roughly the height where particles would begin to get exposed to SCRs as discussed above. We cap it at 0.05 AU arbitrarily. Following the assumption that the wind diffuses quickly after launch (Shang 1998; Hu 2010), subsequent kinematics of solids are calculated without concern as to how the very low-density gas in the wind affects their trajectories.

As the properties of winds are uncertain, in our simple model, we assume the wind accelerates particles to velocities of tens to hundreds of km s$^{-1}$ at launch (Shu, et al. 1996; Liffman 2005; Salmeron & Ireland 2012). Some particles with the right combinations of launch angle and initial velocity reenter the disk, where they mix with local dust and can be later incorporated into planetesimals that form at these greater distances from the Sun. It is during this above-the-disk transport that they would be exposed to a high flux of SCRs and develop their inventory of volatiles such as $^{21}Ne_{cos}$, which is retained due to the low temperatures present in these regions of the disk. The detailed ejection scenario is outlined in Fig. 3.

As illustrated in the Fig. 3a, when a particle moves in the Solar System with a position of (x, y, z), the distance between it and the protosun would be $l = \sqrt{x^2 + y^2 + z^2}$; for the elevation angle $\theta$ and the azimuth angle $\varphi$ we have $sin\theta = z/l, cos\theta = \sqrt{x^2 + y^2}/l, sin\varphi = y/\sqrt{x^2 + y^2}, cos\varphi = x/\sqrt{x^2 + y^2}$ where $\theta > 0$ indicates the angle above the disk midplane and $\varphi = 0$ is aligned with the positive x axis. In Fig. 3b, we put the grain at the yz plane initially for simplicity. Here ($V_y$, $V_z$) is the launch velocity with $Vy=V_{launch}×cos(\alpha_0)$ and $V_z=V_{launch}×sin(\alpha_0)$, while $\alpha_0$ is the initial inclination defined relative to the y-axis. $V_x$ is the initial orbital velocity in the normal direction to the yz plane. The movement of the grain is controlled by the gravity of the protosun as:

$$\frac{d^2z}{dt^2} = -\frac{GM}{l^2}sin\theta \tag{3}$$

$$\frac{d^2y}{dt^2} = -\frac{GM}{l^2}cos\theta sin\varphi \tag{4}$$

$$\frac{d^2x}{dt^2} = -\frac{GM}{l^2}cos\theta cos\varphi \tag{5}$$

Here $G$ is the gravitational constant, $M$ is the mass of the protosun, and $t$ is time. We ignore the gravitational effects of disk.

As for the launch velocity, Shu et al. (1997) proposed a velocity of 140 to 240 km/s for the X-wind. Lee et al. (2017) observed that a protostellar jet can move faster than 100 km s$^{-1}$. Observations of outflows of molecular gas suggested a low velocity < 20 km s$^{-1}$ (Greenhill et al. 2013; Bjerkeli et al. 2016). As stated above, we are agnostic regarding the source of the winds, and thus, we consider $V_{launch}$ ranging from 0 to 200 km s$^{-1}$ and the Keplerian velocity as the initial $V_x$. As for the launch inclination, $\alpha_0$, as we expect the winds to move upward, away from the disk, it should be greater than the elevation angle, $\theta$, and less than π/2. Particles are either accreted onto the disk or expelled from the Solar System. As most refractory grains are found in carbonaceous meteoritical materials that are believed to have formed outside the orbit of proto-Jupiter (Warren 2011; Kruijer et al. 2020), we identify the region of interest for launched particles as those that re-enter the disk at $r > 3$ AU

(Desch, et al. 2018; Weiss & Bottke 2021). Equations (3-5) are solved with the ode45 function in MATLAB with adaptive step size.

**3.3. The Thermal History and the Constraint on the 'Above the Disk' Model**

The thermal history of refractory grains can be divided into two distinct stages. Firstly, upon ejection from the disk, these grains encounter a near-vacuum environment where their thermal energy balance is regulated by solar radiation input and the black-body radiation they emit. Secondly, upon re-entry into the disk, these minerals may experience rapid heating due to gas friction until their relative velocity with respect to the gas decreases significantly. Thus, we consider the thermal history of these grains as being comprised by these two stages.

*3.3.1. Thermal History above the Disk*

When solids move above the disk, solar radiance is the dominant heating source. Solar radiative heating of bodies close to the Sun has been studied for zodiacal dust (Gustafson 1994). The temperature depends on the heliocentric distance, the body size and shape, and the albedo. In calculating the temperature of the grains in our model, we have to make one additional consideration — the transmissivity of refractory minerals. That means, the light absorption efficiency is not 100%. Our model treats solids as spherical and isothermal graybodies (see Fig. 4).

The energy input per unit area can be written as $(1-A)I(r_E/r)^2$. $A$ is the hemispherical reflectance, $I$ is the solar radiance, $r_E$ is the Sun-Earth distance, and $r$ is the heliocentric distance of the particle. Evolutionary tracks in the Hertzsprung-Russell diagram show that the luminosity of pre-main-sequence stars with 1 $M_\odot$ is ~6 times higher than the main-sequence level at the beginning (Palla & Stahler 1993). Therefore, we take the approximate maximum $I = 6I_\odot$ in our calculation to represent the early solar radiance, where $I_\odot$ is the present-day solar radiance (Coddington et al. 2015). The ratio of the emitted energy over the input is given by the Beer–Lambert law, $e^{-\tau \cdot l}$. Here $\tau$ is attenuation coefficient, $l$ is the optical path length equal to $2R cos\theta sin\varphi$, $R$ is the radius of the particle, $\theta$ and $\varphi$ are polar and azimuthal angles, ranging from $-\pi/2$ to $\pi/2$ and from 0 to $\pi$, respectively. The small area unit of the hemisphere, $ds$, is $R^2 cos\theta d\theta d\varphi$, and its projection on the midplane is $ds cos\theta sin\varphi$. The thermal balance can be expressed as:

$$\int_{250\,nm}^{2500\,nm} \int_{-\pi/2}^{\pi/2} \int_0^\pi (1-A)I\left(1-e^{-\tau \cdot l}\right)\left(\frac{r_E}{r}\right)^2 R^2 \cos^2\theta \sin\varphi \, d\lambda \, d\theta \, d\varphi = 4\pi R^2 \varepsilon \sigma T^4 \quad (6)$$

The left-hand side of the equation represents the absorbed energy, and right-hand side is the energy output by thermal radiation, $\lambda$ is the wavelength, $\varepsilon$ is the the infrared emissivity, $\sigma$ is the Stefan-Boltzmann constant, and $T$ is the temperature. $A$, $I$, and $\tau$ are all functions of the wavelength $\lambda$. The solar irradiance is mostly comprised of the shortwave band from 250 nm to 2500 nm today, the range of $\lambda$ in our model.

The reflectance of thousands of materials has been acquired with the Advanced Spaceborne Thermal Emission Reflection Radiometer (ASTER) on NASA's Terra platform (Baldridge et al. 2009). However, the hemispherical reflectance (also known as black sky albedo) of the minerals of interest here have not been determined, and we use other refractory minerals—corundum and forsterite—as analogs of hibonite and spinel. The attenuation coefficient, $\tau$, of corundum and forsterite have been obtained with spectroscopy (Rossman 2010). The emissivity, $\varepsilon$, of refractory minerals ranges from 0.7 to 1.0 for μm-sized grains, and we take an approximate mean of 0.9 (Christensen et al. 2000). The emissivity varies with temperature but would not change too much when heated to hundreds of K, which would not significantly affect our results.

We take the optical parameters ($A$ and $\tau$) from Baldridge et al. (2009) and Rossman (2010) that vary a little in measurements possibly due to the sample heterogeneity. Thus, the results show temperate ranges for grains at different heliocentric distances, as plotted in Fig. 5. The highest temperature experienced by refractory minerals is < 570 K at the beginning of the movement. With such a temperature, the refractory mineral with a typical radius of 50 μm would retain more than 95% of its Ne after $10^5$ years. That is much longer than the travel time above the disk (see section 4.1), meaning that the thermal loss during the transport above the disk is neglectable, and thus we can consider particles accumulating and retaining $^{21}Ne_{cos}$ during transport.

### 3.3.2. Thermal History During Reentering the Disk

As particles re-enter the disk, they will exchange energy and momentum with the surrounding gas. This exchange will lead to the particles warming as they slow down and dynamically equilibrate within their new environments. If particles enter at too high of a velocity, particles may be destroyed or warmed to the point where SCR-produced volatiles could be lost. To estimate what velocities would allow particles to stay at low enough temperatures to retain the implanted noble gases, we model these exchanges following the equations used to describe the evolution of particles in nebular shock waves from previous studies that investigated chondrule formation (Hood & Horanyi 1991; Iida et al. 2001; Ciesla & Hood 2002; Desch & Connolly Jr 2002). The detailed thermal evolution will vary with each particle, depending on the details of the entry location and geometry. Rather than model this for every single particle, we calculated the thermal evolution of particles assuming vertical entry into the protoplanetary disk at 4 AU, described above, to provide a rough estimate. We assume a vertically isothermal temperature of the gas of 350 K, and a surface density of ~3,000 g cm$^{-2}$. The high surface density here would be in the range of what would be expected for a very young, massive solar nebula at the time of CAI formation. We find that particles entering the disk with a velocity of ~30 km s$^{-1}$ reach peak temperatures of nearly 870 K, and the heating lasts for a maximum of a few days. This would lead to a ~20% loss of trapped He via diffusion in spinels with radii of tens of µm (see section 2). Due to a higher retentivity of Ne compared to He, we thus take this as a conservative estimate of the upper limit of the entry velocity particles can have to retain Ne.

## 4. RESULTS

### 4.1. Solids Trajectory Outcomes

In total, we simulated the launch of one million particles, spanning the launch parameters outlined above. Typical trajectories of particles are shown in Fig. 6. Though the detailed movement is controlled by the initial conditions – the combination of starting velocity, initial height, and launch angle – we find that there are three categories of dynamic outcomes: (1) particles with high velocities and launching angles were more likely to be ejected from the Solar System; (2) those with low velocities often fell back onto the inner disk and would be mostly accreted to the Sun (Ciesla 2010b), and (3) a small fraction of the solids (~19% of those simulated here) reentered the solar nebula in a region where they could be accreted by carbonaceous chondrite parent bodies in the outer solar system's disk. Grains launched with an initial velocity range of 81–98 km s$^{-1}$ and a launch angle of 23–87° were more likely to reenter the outer solar nebula with velocities <30 km s$^{-1}$.

To begin, we randomly explored the parameter space outlined above for particle launching, and then ran the model with a high number of particles to sample the parameter space for the third type of particle outcome described above. In total ~55% of these released particles accreted back onto the outer solar System, and about 57% of these grains reentered with speeds less than 30 km s$^{-1}$.

**4.2. Exposure to the Early SCRs.**

When travelling above the disk, the particles get exposed to SCRs directly allowing $^{21}$Ne$_{cos}$ to accumulate. For those reaccreted particles within our focused parameter space, the total exposure time distribution is shown in Fig. 7a&b. The range of time solids spent above the disk spanned <1 to ~32 years with an average of 6.3 years. Given that the SCR flux decreases proportionally to $r^2$, with $r$ being the heliocentric distance, to combine the relationship between the exposure time and irradiation levels seen by the particles, we normalize the exposure time to what it would be at 1 AU to obtain the same integrated fluence. We refer to this as the "effective exposure". The cosmogenic nuclide concentration in the refractory minerals in our model is given by: $s_{msr} = \sum P(r)\Delta t$. Where $s_{msr}$ (cm$^3$ g$^{-1}$ STP) is the concentration of SCR-produced nuclides measured in minerals, $P(r)$ (cm$^3$ g$^{-1}$ Ma$^{-1}$) is the production rate that decreases with $r^{-2}$, and $\Delta t$ is the time step. The equation can be rewritten as:

$$s_{msr} = \sum P(1\,\text{AU})_{msr} \left(\frac{r}{1\,\text{AU}}\right)^{-2} \Delta t = P\,(1\,\text{AU})_{msr} \sum \left(\frac{r}{1\,\text{AU}}\right)^{-2} \Delta t \qquad (7)$$

Where $\sum \left(\frac{r}{1\,\text{AU}}\right)^{-2} \Delta t$ is the effective exposure time, and $P(1\,\text{AU})_{msr}$ is the production rate at $r$ = 1 AU recorded by $^{21}$Ne$_{cos}$. The effective exposure time of reaccreted grains is shown in Fig. 7c&d. The effective exposure times for all particles are between 0.6 to 1 year with an average of 0.76 years. After reentering the disk, the motions of dust are dominated by diffusion and gravitational settling. Consequently, most dust accreted into the protosun, especially those reaccreted at small heliocentric distances (Ciesla 2010a).

**4.3. Production Rate and SCR Flux Calculations**

With the effective exposure time calculated above, we can determine the production rate of $^{21}$Ne$_{cos}$ in the transported refractory minerals. Because the production rate is proportional to the SCR flux, by comparing the production rate at early times to the contemporary one, it is feasible to assess the relative magnitude of the early SCR flux compared to the present-day level. The contemporary production rate of $^{21}$Ne$_{cos}$ can be determined following Trappitsch & Leya (2013) for individual exposed free-floating grains. We apply the chemical composition of refractory minerals to the excel file provided by Trappitsch & Leya (2013) and gain the $P(current, 1\,AU)$ of $^{21}Ne_{cos}$ in the current Solar System. Thus, we can determine the SCR flux needed to match the production rates recorded by the particles using:

$$\frac{P(early, 1\,AU)}{P(current, 1\,AU)} = \frac{J_{early}}{J_{current}} \qquad (8)$$

Here $P(early, 1\,AU)$ equals to $P(1\,AU)_{msr}$ mentioned in the section 4.2, $J_{early}$ is the solar flux intensity in the early Solar System, and $J_{current}$ is the contemporary level.

With mass spectrometry Kööp et al. (2018) determined the SCR-produced $^{21}Ne_{cos}$ concentration in 33 individual refractory grains-hibonites (18) and spinel-bearing inclusions (hereafter spinel for simplicity) (15), and found that at least 78% of the hibonites were exposed to the protosun before the incorporation and that only ~7% of spinels record the exposure. The $^{21}Ne_{cos}$ abundances range from 1.02±0.17 to 29.89±0.70 ($10^{-8}$ cm$^3$ g$^{-1}$ STP) with a weighted mean of 5.73±0.02 ($10^{-8}$ cm$^3$ g$^{-1}$ STP) in hibonites and from 0.05±0.01 to 7.90±0.14 ($10^{-8}$ cm$^3$ g$^{-1}$ STP) with a weighted mean of 1.00±0.01 ($10^{-8}$ cm$^3$ g$^{-1}$ STP) in spinels (see data in Supplementary Table 4 in Kööp et al. (2018)). After subtracting the parent body exposure (1.6 Ma for Murchison), the weighted mean of $^{21}Ne_{cos}$ is 5.63±0.02 and 0.62±0.01 ($10^{-8}$ cm$^3$ g$^{-1}$ STP) for hibonites and spinels, respectively.

Combining the information above, we calculate the average early Solar System $^{21}Ne_{cos}$ production rate at $r = 1$ AU, that is, 7.4×10$^6$ ($10^{-8}$ cm$^3$ g$^{-1}$ Ma$^{-1}$) for hibonites during exposure and 8.1×10$^5$ ($10^{-8}$ cm$^3$ g$^{-1}$ Ma$^{-1}$) for spinels during exposure. The contemporary $^{21}Ne_{cos}$ production rate for a floating grain at the same position calculated by Trappitsch and Leya (2013) is 1.06 ($10^{-8}$ cm$^3$ g$^{-1}$ Ma$^{-1}$) and 2.08 ($10^{-8}$ cm$^3$ g$^{-1}$ Ma$^{-1}$) for hibonites and spinels, respectively. By comparing the production rates above (see equation 8), we obtain a $10^{6.8}$ times higher SCR flux during the hibonite transport and a $10^{5.6}$ times higher flux during the spinel transport compared to the contemporary level.

## 5. DISCUSSION

### 5.1. Robustness of the Model and Uncertainty Estimate

In the wind-ejection model we assume solids are launched from a region close to the young Sun and fall back onto the disk in the outer Solar System, their initial speed and launch angle are constrained as described in section 4.1.

Consequently, when we input all possible configurations of the initial conditions to the model, the resulted effective exposure times have a narrow range between 0.6 and 1 year with an average of 0.76 years. The nominal uncertainty from the exposure time estimate in this model is less than one third. When running the model with different initial parameters of disk structure as defined by temperature ($T_0$=700 to 1000 K), surface density ($\Sigma_0$=3000 to 12000 g cm$^{-1}$), and CAI launch region (0.08 to 0.12 AU), the average effective exposure time varied within <10%.

The second uncertainty is that we assume perfect retention of all $^{21}$Ne, which would not be the case if the particles experience significant heating after exposure. In this work, the treatment of reentry heating is simplified. We set 30 km s$^{-1}$ as the threshold velocity above which particles would lose their volatiles and below which particles would keep their full noble gas inventories. Obviously, the heating effect is not a single-step function and diffusion for any given time and temperature will result in partial volatile loss via the thermal diffusion as discussed in section 2. For example, if we set 50 km s$^{-1}$ as the threshold of reentry velocity, the grains would be heated up to ~1070 K, ~50% of He would be lost in spinel group minerals, and <50% of He would be lost in hibonites. Given Ne is more retentive than He, its thermal loss would be much less (section 3.3.2). Also, the average effective exposure time decreases to 0.72 year. Thus, the uncertainty arising from thermal loss is much less than 50%.

The third uncertainty is from analyses of $^{21}$Ne$_{cos}$ in hibonites and spinels that involves the grain volume and mass estimate, mass spectrometric blank correction, calibration, and the subtraction for the atmospheric Ne component. Previous work (Heck et al. 2009; Kööp, et al. 2018) have identified the biggest uncertainty source as the volume and mass estimates, which can be overestimated by a factor of 1.6 for grains with irregular shapes. Therefore, the SCR flux could be underestimated by the same factor at most.

Combing all the three uncertainties, the early SCR flux could be overestimated by a factor of 1.3 or be underestimated by as much as a factor of 4.0, meaning the enhancement compared to the present-day level is between $10^{6.7}$ and $10^{7.4}$ during hibonite formation. For the irradiation record in spinels, only ~7% of them (1/15) show pre-compaction irradiation, and the irradiated spinel's Ne isotopic composition is indistinguishable from GCR-Ne. Both features are similar to other irradiated components in Murchison meteorites that can be caused by the mixing of surficial and deeper materials in the regolith (Caffee et al. 1987; Hohenberg et al. 1990; Roth, et al. 2011; Riebe et al. 2017). This implies that, in contrast to hibonites, spinels may have never been exposed to a higher flux of SCRs as individual grains in space. The low fraction of spinel grains with $^{21}Ne_{cos}$ excess were either residing in the parent body regolith before compaction, where they were exposed to GCRs for tens of Myr or, alternatively, they were exposed at the surface to an enhanced SCR flux for <1 Myr.

The reasons why most spinel grains may not have been exposed to an early enhanced SCR flux could be due to (1) a later formation of spinels, after the SCR flux decreased. It was suggested that in the solar nebula hibonites formed before spinels based on the larger isotopic anomalies and higher condensation temperatures of hibonites (Liu et al. 2012; Davis & Richter 2014; Kööp et al. 2016a; Kööp et al. 2016b). (2) different disk dynamics and conditions that did not lead to ejection of spinels above the solar nebula disk that prevented exposure to SCRs. Therefore, we provide a value of a nominally enhanced SCR flux of $10^{5.6}$ inferred from spinel cosmogenic Ne data only for reference here (section 4.3).

Other systematic uncertainties of our model include, but are not limited to, the structure of the disk, the scattered SCRs in the disk, the efficiency and grain-size dependence of the particle launch mechanism, shielding by nebular gas and dust, and the energy spectrum of early SCRs. A detailed treatment of these are beyond the scope of this study.

**5.2. Comparison to the Astronomical Observations**

The ages of refractory minerals relative to astronomical chronology are not well anchored, but their isotopic composition (O, Al-Mg, etc.) and the physical model of molecular cloud collapse suggest that most of them could have formed before the Class II phase of YSOs (Liu et al. 2019; Brennecka et al. 2020), which corresponds to the < first Myr of YSOs evolution. Especially, hibonites, displaying notable isotopic anomalies in $^{48}Ca$ and $^{50}Ti$ and lacking radiogenic $^{26}Mg$ excess from extinct radioactive $^{26}Al$, are believed to precede most other refractory minerals (Kööp, et al. 2016a; Krot et al. 2020). Thus, to interpret the record of SCR-produced nuclides in these minerals we assume they were present and exposed to SCRs during the YSO phase of the Sun.

Our knowledge of YSOs activities and related proton fluxes is based on the observation of stellar flare-generated X-ray emission. Feigelson, et al. (2002) studied 43 solar-mass YSOs in the Orion Nebula Cluster (ONC) and used the "characteristic" X-ray emission around $\log(L_X)$ ~30 erg s$^{-1}$ to infer a minimum of an $10^5$-fold enhancement in energetic protons for the young Sun. Studies of the young stellar cluster NGC 602a in the Small Magellanic Cloud (SMC) and intermediate-mass (2–5 $M_\odot$) pre-main-sequence stars also found high X-ray luminosities that were attributed to the high activity of the YSOs (Oskinova et al. 2013; Nuñez et al. 2021). Our findings suggest a higher activity for the young Sun than that inferred from astronomical observations of possible analogs (Feigelson, et al. 2002; Wolk, et al. 2005). We attribute this to the following. (1) Besides the "characteristic" X-ray emission previously used to explain the record in meteoritic refractory minerals of the early SCR flux, superflares and megaflares with higher X-ray emissions from YSOs have been found in the past twenty years and were systematically discussed recently (Getman & Feigelson 2021; Getman et al. 2021). Though the scaling of X-ray emission for superflares and megaflares to energetic proton flux is not well-constrained, such flares would certainly increase charged particle flux. (2) Here we try to estimate the early SCR flux using the method from Feigelson et al. (2002). The most luminous X-ray solar flares in recent solar cycles had log $L_X$ (peak) of ~ 28.5 ergs s$^{-1}$ and a total energy log ($E_X$) of ~32.5 ergs. Such strong solar flares occur roughly once a year (Sammis et al. 2000; Feigelson, et al. 2002). Superflares and megaflares from YSOs have peak X-ray luminosities of log ($L_X$) = 30.5–34.0 ergs s$^{-1}$ and X-ray energy log ($E_X$) of 34–38 ergs. The occurrence rage of flares with log ($E_X$) ~ 32.7 erg for a low mass YSO with an age < 1 Myr are extrapolated to occur ~$10^6$ times per year (Getman & Feigelson 2021). Thus, the total flare X-ray emission is >$10^6$ times higher than the contemporary level. (Feigelson et al. 2002) used $10^4$ enhancement in X-ray emission to extrapolate a factor of $10^5$ in SCR flux. If we take the same consideration, $10^6$-enhanced X-ray emission would lead to a >$10^7$ times higher SCR flux in the early Solar System compared to the present-day level.

**5.3. Interpretation of the Cosmogenic Be Record in Early Solar-System Minerals**

The SCR flux from the young Sun would have had important implications for the evolution of protoplanetary disks, the cloud of gas and dust from which the Sun and the rest of the stellar system formed. For example, energetic particles ionize atoms and molecules in the disk (Rab et al. 2017), creating charged particles and changing the disk chemistry (Henning & Semenov 2013), and heat materials they encounter, etc. Here we focus on how our new SCR flux estimate based on Ne can be used to interpret data from $^{10}$Be, another cosmogenic nuclide analyzed in CAIs. While SCRs from the young Sun produced many other cosmogenic nuclides that are preserved in CAIs, including short-lived nuclides (SLRs) such as $^{26}$Al, $^{36}$Cl, $^{41}$Ca, and $^{60}$Fe, they may only be responsible for a small fraction of the inventories of some of these nuclides whereas the bulk of most is produced by stellar nucleosynthesis (Goswami et al. 2001). Among cosmogenic nuclides, $^{10}$Be, along with $^{6}$Li, is unique because it is formed only by spallation reactions and not by stellar nucleosynthesis (McKeegan et al. 2000; MacPherson et al. 2003; Clayton 2003; Chaussidon & Gounelle 2006; Bricker and Caffee 2010, Liu, et al. 2010; Srinivasan & Chaussidon 2013; Sossi, et al. 2017). Liu, et al. (2010) discussed two possible irradiation scenarios to explain $^{10}$Be in CM hibonites: irradiation of hibonite solids and irradiation of solar gas. In our model, the proton fluence hibonite was exposed to above the disk only accounts for only ~2% of the $^{10}$Be production in hibonites, and the majority should have been acquired before. Given the effective shielding from high-energetic particles by the nebula, grains within the disk are not likely to be able to receive much irradiation (Trappitsch & Ciesla 2015). Thus, most $^{10}$Be nuclides in hibonites were inherited from the nebula and incorporated during condensation. Possible origins are SCR-induced spallation of C and O of nebular gas (Jacquet 2019) or spalllation reactions in the proto-solar atmosphere (Bricker and Caffee 2010). This also puts a constraint on the difference of $^{10}$Be in hibonites from CM chondrites and CAIs from CV chondrites. The inferred initial $^{10}$Be/$^{9}$Be ratio of ~5×10$^{-4}$ in CM chondrite hibonites is lower than the best-constrained $^{10}$Be/$^{9}$Be of ~7×10$^{-4}$ in CV CAIs (Liu, et al. 2010; Dunham et al. 2022). Given the high SCR flux, the production rate of $^{10}$Be in the nebula would have been ~100 times higher than determined before and largely surpassed the decay rate of $^{10}$Be ($t_{1/2}$= 1.4 Myr) (Jacquet 2019). The difference of the $^{10}$Be/$^{9}$Be ratios between CM hibonites and CV CAIs can be explained by an earlier formation of CM hibonites compared to CV CAIs, with a time difference Δt << $t_{1/2}$ of $^{10}$Be. The cosmogenic $^{10}$Be produced during Δt was incorporated into CV CAIs when they formed which resulted in higher $^{10}$Be/$^{9}$Be ratios compared to the ones measured in CM hibonites. Because only a small number of CM hibonites have been analyzed for $^{10}$Be (Liu et al. 2010) other possibilities, such as heterogeneity of $^{10}$Be distribution in the early solar nebula disk during CM hibonites formation cannot be excluded.

## 6. CONCLUSION

(1) In this work, we modeled the outward movement of the refractory oxides hibonite and spinel from CM chondrite Murchison in the early Solar System with two approaches and inferred their effective exposure times to SCRs. In the diffusion model, most particles were largely restricted to regions of the disk where primary SCRs could not penetrate and did not get exposed. In the wind-driven model, the effective exposure time for those grains that were reincorporated into the solar nebula was ~0.76 years on average.

(2) With the modeled exposure times and the measured $^{21}Ne_{cos}$ concentration in hibonites and spinels from Murchison we provide a new estimate of the SCR flux of the young Sun during the hibonite formation period. With the average effective exposure time, the enhanced SCR fluxes are estimated to have been ~$10^7$ times higher during hibonite exposure than the current levels. The irradiation record of spinels is consistent with exposure at the parent body surface or in the regolith. Thus, spinels may have never experienced the high SCRs emitted from the young Sun as individual grains in space.

(3) Our findings imply that only a few percent cosmogenic $^{10}Be$ in hibonite from Murchison was produced during SCR-exposure of solids, while most of it was inherited from the gas phase, possibly from the interactions between SCRs and disk.

(4) The flare activity of the young Sun derived here is largely consistent with the astronomical observations of flares of low- to intermediate-mass YSOs in their first Myr.

## Acknowledgements:
PRH and XY acknowledge funding through NASA Emerging Worlds and FINESST grants (80NSSC21K0389 to PRH and 22-PLANET22-0160 to PRH and XY), the Field Museum Science Innovation Award, and the TAWANI Foundation, and appreciate the helpful discussions with Andrew M. Davis, Marc W. Caffee, and Steven J. Desch. Comments from an anonymous reviewer helped improve the manuscript.

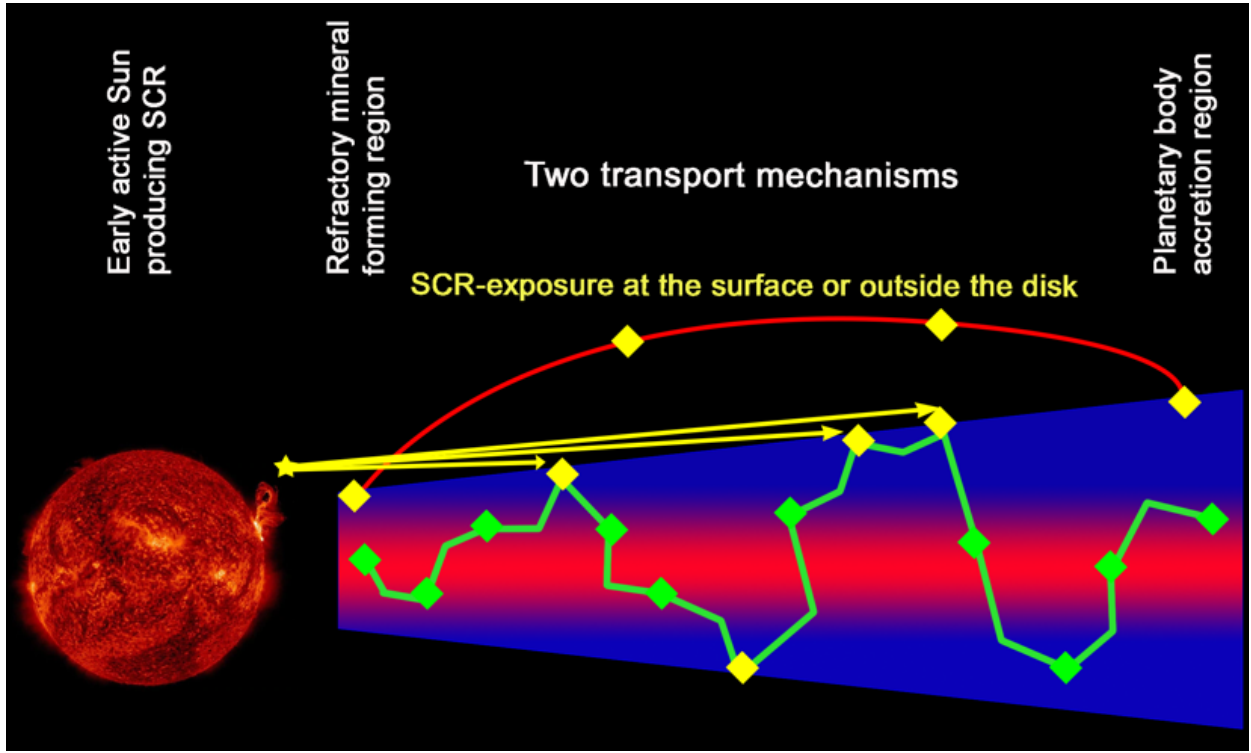

Fig. 1. Conceptual schematic of trajectories of refractory minerals shortly after they formed close to the Sun ~4.6 Gyrs ago. Yellow diamonds indicate where SCR-irradiation occurred. The red curve represents the trajectory of a particle accelerated by the X-wind. The green curve represents the diffusive movement of a particle within the disk which is shown in schematic cross section with a trapezium shape. Chromosphere image of the Sun courtesy of NASA.

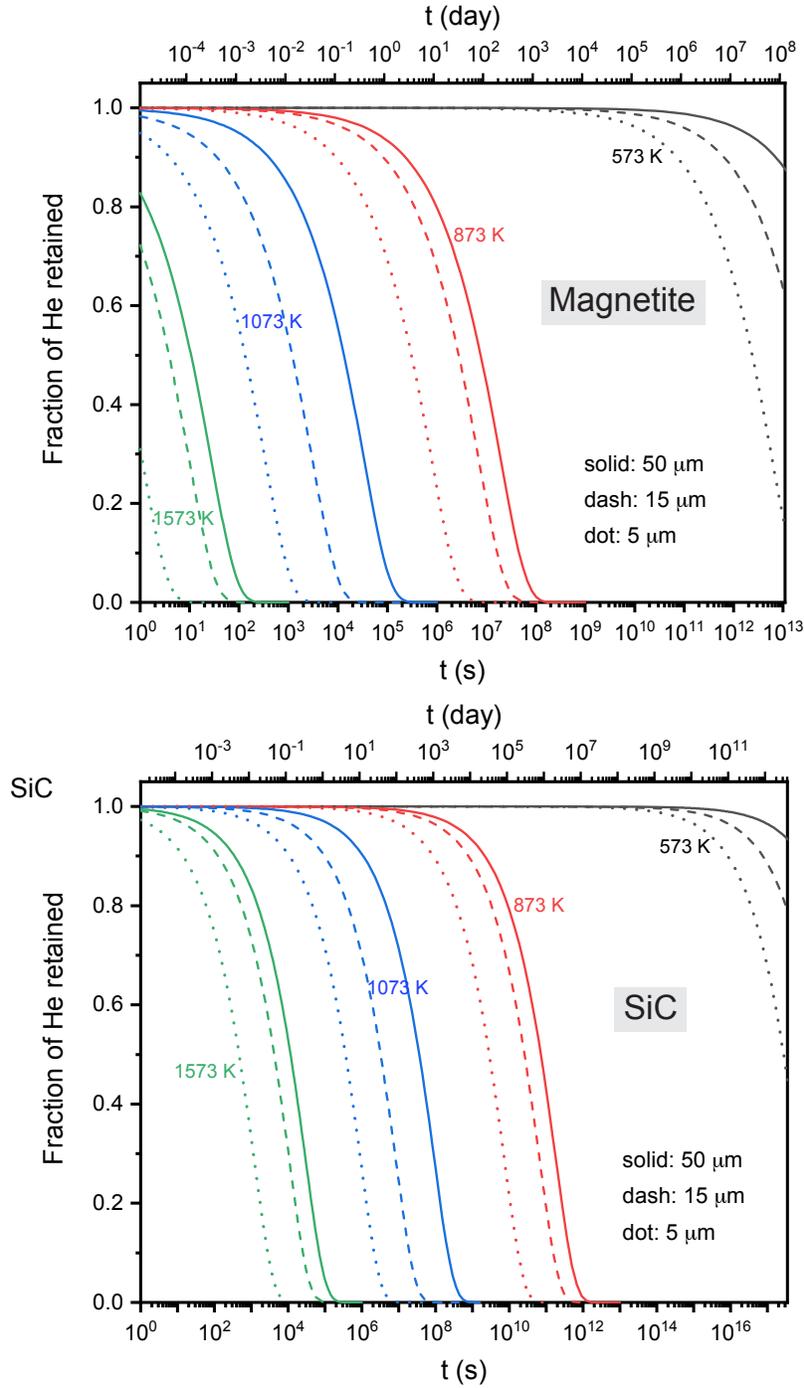

Fig. 2. Fraction of He remaining during isothermal heating at 300 °C, 600 °C, 900 °C, and 1300 °C as a function of time from spherical grains of magnetite and SiC with radii of 5, 15, and 50 μm. Curves are plotted using the diffusion parameters from Blackburn, et al. (2007) and Cherniak, et al. (2016).

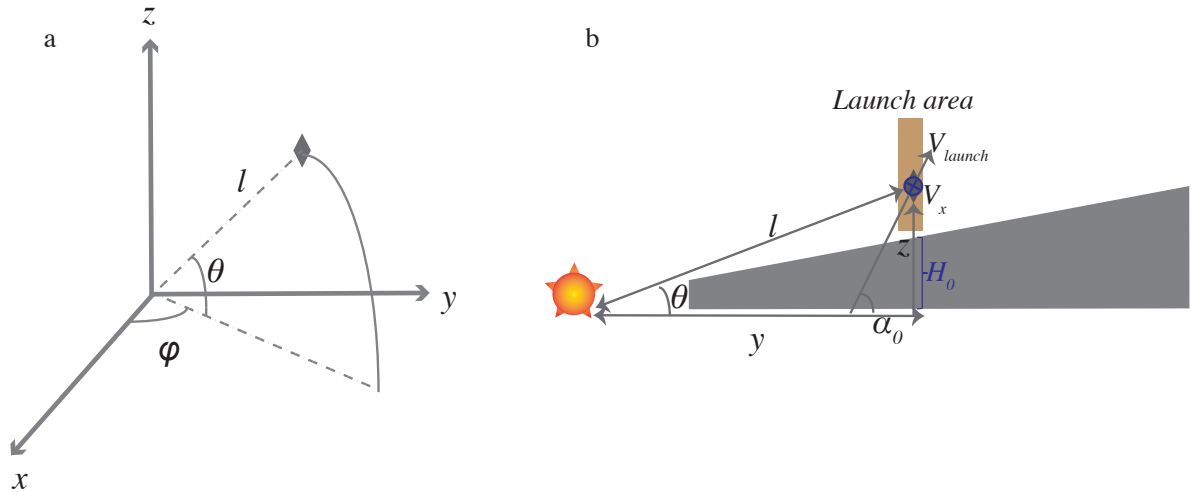

Fig. 3. Geometry of our simple "above the disk" model. (a) A particle (diamond symbol) in the Solar System whose coordinate is (x, y, z) and the protosun at the origin. (b) The schematic displays the yz plane of an early protoplanetary disk where the disk is truncated close to the protosun. The wind launches solids (diamond symbol) with initial velocities of $V_{launch}$ in this cross section and of $V_x$ in the normal direction. In the approximation of a geometrically thin disk, we assume that the gravitational force from the protosun is the only force on the particle after launch, and that we neglect all others after launch. The path of the grain follows a conic section.

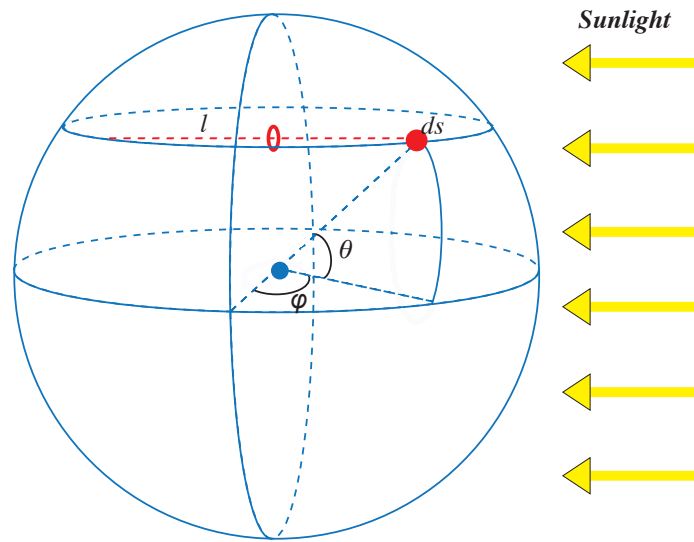

Fig. 4. Sketch of an irradiated spherical grain. Sun light (yellow arrows) is reflected at the surface and pass through the grain with some energy absorbed. The red dashed line indicates the optical path. The red filled symbol is a small area unit of the surface, and the red open one is its projection on the midplane.

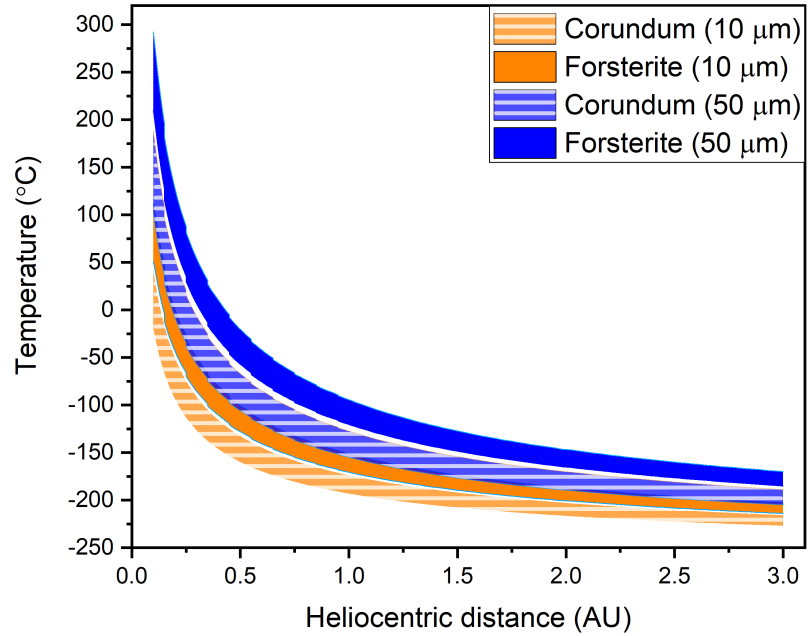

Fig. 5. Temperature of grains with different mineralogy vs. heliocentric distance. Two types of minerals are indicated by colors. The shape of minerals is sphere with radii of 10 and 50 μm. Grains move from their launching position (0.1 AU) to 3 AU.

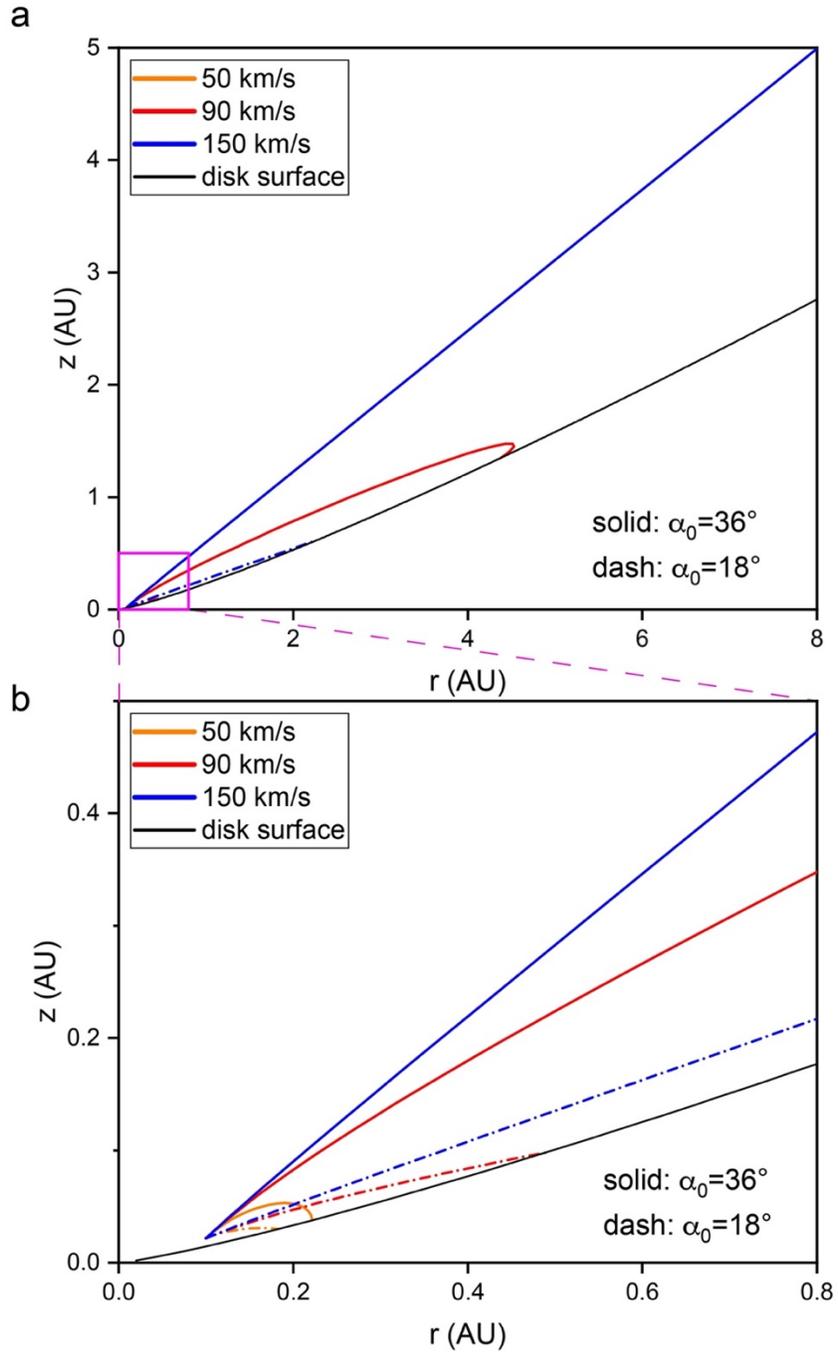

Fig. 6. Representative trajectories of solids with initial launching velocities of 50, 90, and 150 km s$^{-1}$ and with launching angles of $\pi/5$ and $\pi/10$. **a** Radial distance of 0–8 AU; **b** Inset at r=0–0.8 AU. Solids with the highest launching angles ($\pi/5$) and highest velocities (150 km s$^{-1}$) are ejected while those with $V_{launch}$ < 50 km s$^{-1}$ accreted close to the Sun.

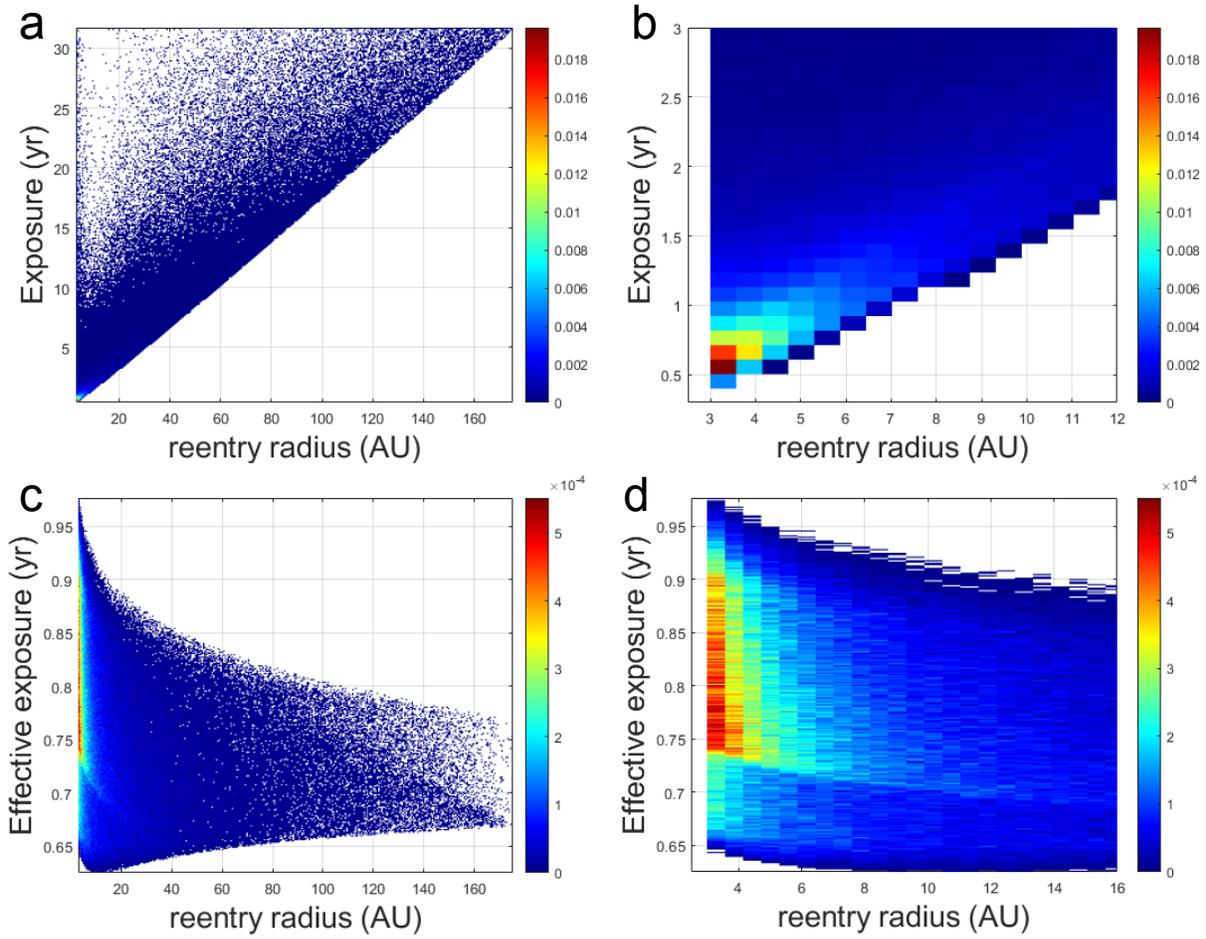

Fig. 7. Exposure time vs. reentry radial distance of grains reaccreted onto the outer Solar System with reentry velocities less than 30 km s$^{-1}$. **a** is the total traveling time above the disk, **b** is a magnified view of the lower-left region of **a**. **c** is the effective exposure time. **d** is an enlarged depiction of the left area of **c**. Color represents the normalized number density.